\documentclass[12pt]{article}

\input epsf

\def\d{\mbox{d}}

\begin{document}

\title{Unstable-particles pair production in MPT \\ approach in
NNLO\thanks{Contribution to the proceedings of {\it 14th Lomonosov
Conference on Elementary Particle Physics}, 19-25 August 2009,
Moscow}}

\author{ M.L.Nekrasov \\
\small Institute for High Energy Physics, 142284 Protvino, Russia
\vspace*{-12mm} }
\date{}

%Talk given at 14th Lomonosov Conference on Elementary Particle Physics, Moscow, 19-25 Aug 2009.

\maketitle

\begin{abstract}
We consider pair production and decay of fundamental unstable
particles in the framework of a modified perturbation theory
treating resonant contributions in the sense of distributions.
Outcomes of a numerical simulation up to the NNLO with taking into
account universal massless-particles contributions are presented.
\end{abstract}

In realistic descriptions of the processes of production and
decays of fundamental unstable particles the gauge cancellations
and high accuracy of the calculation of resonant contributions
must be provided. A modified perturbation theory (MPT)
\cite{Tkach,MPT} is one of the approaches intended for the
solution of these problems. Its main feature is the direct
expansion of the probability instead of the amplitude in powers in
the coupling constant, on the basis of distribution-theory
methods. As the object to be expanded (the cross-section)~is gauge
invariant, the gauge cancellations in the MPT must be
automatically maintained. However the accuracy of the calculation
of resonant contributions remains vague. For clearing up this
question numerical simulations are required.

In this report I present results of calculations in the MPT up to
the NNLO in a model defined as the improved Born approximation for
the process $e^{+} e^{-} \to \gamma,Z \to t\bar t \to W^{+}
b\:W^{-}\bar b$. The total cross-section of the process is
\begin{equation}\label{not1}
\sigma (s) =
 \int\limits_{s_{\mbox{\tiny min}}}^s \frac{\d s'}{s}\:
 \phi\!\left(\frac{s'}{s};s\right)
\!\!\!
   \int\limits_{\quad
   {\displaystyle\mbox{\scriptsize $s$}}_
                    { i \mbox{\tiny min} } \atop
              \quad {\displaystyle\mbox{\scriptsize $s$}}_
                    { 2 \mbox{\tiny min} }}
            ^{\!\!\infty}   \!\!\!\!\!\!\!\!
   \int\limits^{\infty} \d s_1 \, \d s_2 \;\;
   \hat\sigma(s'\,;s_1,s_2) \left(1\!+\!\delta_{c}\right),
\vspace*{-0.5\baselineskip}
\end{equation}
where $\phi(s'/s;s)$ is the flux function,
$\hat\sigma(s\,;s_1,s_2)$ is the exclusive cross-section,
$\delta_{c}$ stands for soft massless-particles contributions,
$s_1$ and $s_2$ are virtualities of the top quarks. With massless
$b$ quarks $s_{1\,\mbox{\scriptsize min}} =
s_{2\,\mbox{\scriptsize min}} = M_{W}^2$, $s_{\mbox{\scriptsize
min}}=4M_W^2$. The $\phi$ we consider in the leading log
approximation. By $\delta_{c}$ the Coulomb factor is me\-ant,
taken into consideration in the one-gluon approximation with
conventional resummation, not affecting the Breit-Wigner (BW)
factors (see details in \cite{MPT}).

In $\hat\sigma(s\,;s_1,s_2)$ the kinematic and BW factors can be
extracted,
\begin{equation}\label{not4}
\hat\sigma(s\,;s_1,s_2) = \frac{1}{s^2} \,
  \theta(\sqrt{s}-\!\sqrt{s_1}-\!\sqrt{s_2}\,)
  \sqrt{\lambda(s,s_{1},s_{2})}\;\Phi(s;s_1,s_2)
  \> \rho(s_{1}) \> \rho(s_{2})\,.
\vspace*{-0.3\baselineskip}
\end{equation}
Further, $\Phi(s;s_1,s_2)$ may be determined in the conventional
perturbation theory, but the BW factors $\rho(s_{1}) \rho(s_{2})$
must be determined as asymptotic series in the distributions sense
in powers in the coupling constant $\alpha$. For isolated
$\rho(s)$ in the case of a smooth weight, the series up to the
NNLO looks as follows \cite{Tkach}:
\begin{equation}\label{not7}
 \rho(s) \;=\; \delta(s\!-\!M^2) \, + \,
 PV \;\mbox{Taylor}\bigl\{\rho(s)\bigr\} +
 \sum\limits_{n\,=\,0}^2 c_{n}\, \delta^{(n)}(s\!-\!M^2) +
 O(\alpha^3)\,.
\vspace*{-0.6\baselineskip}
\end{equation}
Here $M$ is the top quark mass, $\delta(\cdots)$ is the
$\delta$-function, $\delta^{(n)}$ is its $n$-th derivative. The
`$PV$ Taylor' means the naive Taylor expansion with
principal-value prescription for the poles. Coefficients $c_n$
within the NNLO are determined by three-loop self-energy
contributions (for explicit expressions see~\cite{MPT}).

\begin{figure}[t]
\hbox{ \hspace*{-3pt}
       \epsfxsize=425pt \epsfbox{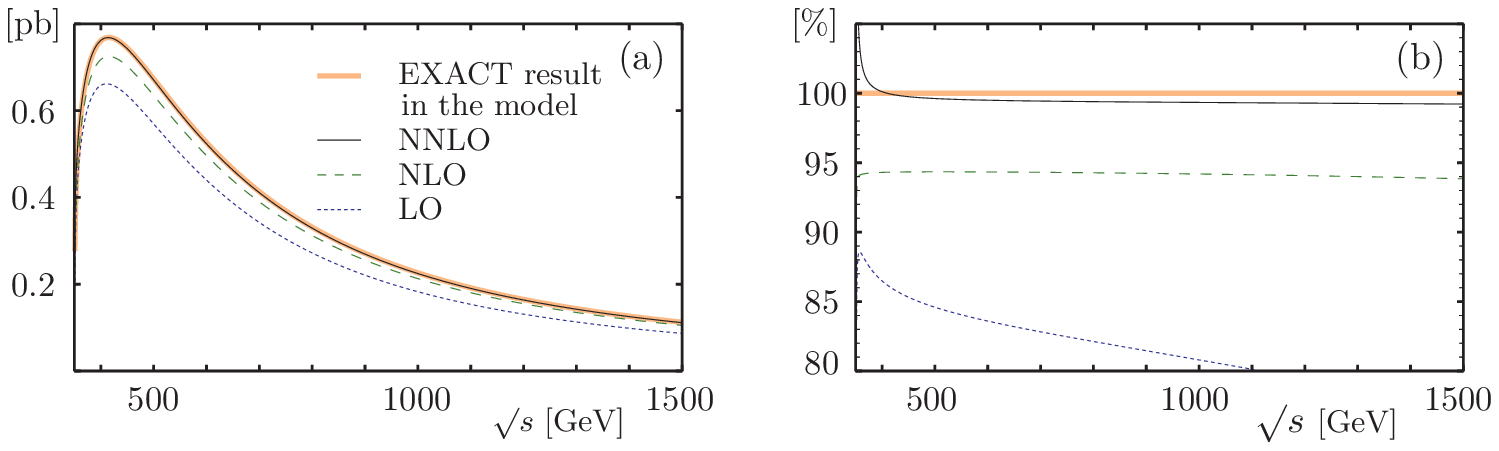}}{\vspace*{-6mm}}
\end{figure}
\begin{table}[t]
\caption{\small The total cross-section in pb and in \% with
respect to the exact result.}
\begin{center}
\begin{tabular}{| c | c c c c | }
\hline\noalign{\medskip} \\[-6mm]
 $\quad \sqrt{s}$ (TeV) $\quad$    & $\quad \sigma \quad$      &
 $\quad \sigma_{LO} \quad$        & $\quad \sigma_{NLO} \quad$ &
 $\quad \sigma_{NNLO}   \quad$ \\
\hline\noalign{\medskip} \\[-6mm]
 0.5                               & 0.6724         &
 0.5687          &  0.6344         & 0.6698(7)          \\
                                   & {\small 100\%} &
 {\small 84.6\%} & {\small 94.3\%} & {\small 99.6(1)\%} \\
\hline\noalign{\medskip} \\[-6mm]
 1                                 & 0.2255         &
 0.1821          &  0.2124         & 0.2240(2)          \\
                                   & {\small 100\%} &
 {\small 80.8\%} & {\small 94.2\%} & {\small 99.3(1)\%} \\[-1mm]
\noalign{\smallskip}\hline
\end{tabular}
\end{center}\vspace*{-5mm}
\end{table}

Unfortunately, the weight in (\ref{not4}) is not smooth because of
the $\theta$-function~and the square root in the kinematic factor.
However it can be analytically regularized by substitution
$[\lambda(s,s_{1},s_{2})]^{1/2} \to
[\lambda(s,s_{1},s_{2})]^{\nu}$. The singular integrals then can
be analytically calculated and after removing the regularization
the results become finite. Moreover, the expansion remains
asymptotic~\cite{MPT}.

The further numerical calculations, I carry out by means of rather
general FORTRAN code with double precision. Parameters of the
model I determine as follows: $M = 175 \: \mbox{GeV}$, $M_W = 80.4
\: \mbox{GeV}$. The results for the total cross-section above the
threshold are shown in the figure (notation is obvious). The
results in numerical form at typical ILC energies are presented in
the table.

It should be emphasized a very stable behavior of the NLO and
NNLO~approximations at the energies above and near the maximum of
the cross-section (this result to a large extent is
model-independent \cite{model}). Near the~maximum the discrepancy
of the NNLO is within $\pm 0.5 \%$, which seemingly is sufficient
for the ILC in the case of the top quarks. The further increase of
the precision is possible at the proceeding to the NNNLO
\cite{MPT}. Another way for boosting~the precision may be based on
the combined use of the MPT, when the loop corrections are handled
in the MPT framework but Born contribution in the conventional
approach with the Dyson-resummation in the unstable-particles
propagators (on the analogy of the actual practice of application
of~DPA).


\begin{thebibliography}{99}
\bibitem{Tkach}
F.V.~Tkachov, in Proc. of the 32nd PNPI Winter School,
St.Petersburg, Ed. by Ya.I.Azimov et al., PNPI, 1999, p. 166
[hep-ph/9802307].
\bibitem{MPT}
M.L.Nekrasov, Int.Mod.Phys. A24 (2009) 6071 [arXiv:0709.3046].
\bibitem{model}
M.L.Nekrasov, arXiv:0912.1025.
\end{thebibliography}
\end{document}